\documentclass[12pt]{article}
\usepackage{amsmath,amssymb,amsthm,amsxtra,overpic,bbm,bm,epsfig,subfigure}
\usepackage[colorlinks=true,citecolor=blue,linkcolor=blue]{hyperref}
\usepackage{mathrsfs}
\usepackage{enumitem}
\usepackage{graphicx}
\usepackage{color}
\usepackage{comment}
\usepackage{epstopdf}
\usepackage{float}
\usepackage{cite}
\usepackage{slashed,stmaryrd}

\def\thefootnote{\fnsymbol{footnote}}

\begin{document}

\vspace{0.2cm}

\begin{center}
{\Large\bf A complete optical model for liquid-scintillator detectors}
\end{center}

\vspace{0.2cm}

\begin{center}
{\bf Yan Zhang $^{a,b}$},
\quad
{\bf Ze-Yuan Yu $^{a}$},
\quad
{\bf Xin-Ying Li $^{a}$}\footnote{Now at Neusoft Medical Systems Co., Ltd.},
\\
{\bf Zi-Yan Deng $^{a}$},
\quad
{\bf Liang-Jian Wen $^{a}$}\footnote{E-mail: wenlj@ihep.ac.cn}
\\
{\small $^a$Institute of High Energy Physics, Chinese Academy of Sciences, Beijing 100049, China}\\
{\small $^b$School of Physical Sciences, University of Chinese Academy of Sciences, Beijing 100049, China}\\
\end{center}

\vspace{1.5cm}

\begin{abstract}
Liquid scintillator (LS) is widely used in various neutrino oscillation experiments, in particular, the reactor neutrino experiments. The complex absorption and re-emission processes of optical photons are known to be an important source of the non-linear and non-uniform response of LS detectors. Precise simulation of light propagation in LS is highly desirable to model the detector response and reduce the systematic errors. In this paper, we develop a novel optical model which can completely deal with the competitive photon absorption and subsequent re-emission processes of the LS components. It allows to directly plug in the laboratory measurements of the LS components to model any LS composition. Extensive measurements have been performed to obtain the essential optical parameters for this model. We validate the model with a bench-top experiment featuring a small LS volume. Furthermore, we demonstrate that for any given detector geometry, this model provides the capability of optimizing the LS recipe to maximize the light collection. It is valuable for designing future LS-based detectors and improving the agreement between Monte Carlo and data for current neutrino experiments.
\end{abstract}

\begin{flushleft}
\hspace{0.9cm} Key words: optical model, liquid scintillator, absorption and re-emission, neutrino experiment
\end{flushleft}

\def\thefootnote{\arabic{footnote}}
\setcounter{footnote}{0}

\newpage

\section{Introduction}
\label{sec:introduction}

The liquid scintillator (LS) technique has been widely used in neutrino experiments for more than sixty years and, in particular, in neutrino oscillation studies with reactors~\cite{Vogel:2015wua}. Nowadays neutrino oscillation experiments have entered an era of precision measurement, where LS-based neutrino detectors such as Daya Bay~\cite{DayaBay:2012aa} and JUNO~\cite{An:2015jdp} will continue to play leading roles.
Recent LS-based neutrino detectors mostly adopt a ternary LS that consists of solvent, primary fluor and wavelength shifter. Table.~\ref{tab:exp} shows the LS composition of several monolithic neutrino detectors. Linear alkylbenzene (LAB) is popularly used as LS solvent. The 2,5-diphenyloxazole (PPO) and p-bis-(o-methylstyryl)-benzene (bis-MSB) are commonly used as primary fluor and wavelength shifter, respectively. Their absorption bands and emission spectra overlap with each other, resulting in a complex sequence of absorption and re-emission for the optical photons from scintillation and \v{C}erenkov processes. Such effects are an important source of the non-uniform and non-linear energy response in a LS detector. Thus, a precise understanding of the photon propagation between the fluorescent molecules will help to reduce the absolute energy scale uncertainty. A sophisticated optical model based on Monte Carlo simulations is the key to understanding the experimental data of the currently running LS-based experiments and guide the design of future LS detectors.

\begin{table}[!htb]
\begin{center}
\caption{LS composition in some recent neutrino experiments.}
\label{tab:exp}
\begin{tabular}{ccccc}
\hline
Detector & Solvent & PPO & bis-MSB & Target size \\ \hline
KamLAND~\cite{Abe:2008aa} & PC + dodecane & 1.36 g/L & -- & $\Phi$(13 m) \\
Borexino~\cite{Alimonti:2008gc} & PC & 1.5 g/L & -- & $\Phi$(8.5 m) \\
Daya Bay~\cite{DayaBay:2012aa} & LAB & 3 g/L & 15 mg/L & $\Phi$(3 m) \\
Double Chooz~\cite{Aberle:2011ar} & n-dodecane + $o$-PXE & 7 g/L & 20 mg/L & $\Phi$(2.3 m) \\
RENO~\cite{Park:2013nsa} & LAB & 3g/L & 30 mg/L & $\Phi$(2.87 m) \\
SNO+~\cite{Andringa:2015tza,Fischer:2018squ} &  LAB  &  2 g/L  &  15 mg/L  &  $\Phi$(12 m) \\
\hline
\end{tabular}
\end{center}
\end{table}

For LS detectors, the light output is one of the driving factors to optimize the LS composition. For medium size LS detectors like Daya Bay, the recipe was optimized by measuring the light output of a small LS cell~\cite{Ding:2008zzb,Ding:2008thesis}, based on the empirical assumption that the absorption and re-emission processes are not important beyond a scale of a few centimeters. However, later on we realized that, in a large-scale LS detector, the effect of absorption and re-emission on light propagation can still be significant at meters scale, comparable to that of Rayleigh scattering. Thus, the optimal LS recipe will depend on the detector size, and it is useful to develop an optical model that can completely describe photon propagation in presence of all LS components. This is particularly true for future large scale experiments like JUNO and similar proposals~\cite{Fischer:2018zsr}. In this paper, we take the ternary LS composed of LAB, PPO and bis-MSB as an example, and introduce a novel optical model to simulate the light propagation in LS detector. Prior to this work, extensive measurements~\cite{Wen2010,Xiao2010,Li2011,Feng:2015tka,Zhou:2015gwa} have been performed to understand the photon transport and extract the optical properties of each LS component.

The paper is organized as follows: the details of the optical model and measurements of the optical properties are described in Section~\ref{sec:opModel}. We present a bench-top experiment with a small LS volume to validate the model and show good agreement between the data and the Monte Carlo prediction in Section~\ref{sec:experiment}. Moreover, we demonstrate the possible application of our optical model to optimize the design of future LS detectors in Section~\ref{sec:modelApp}. Finally, we summarize our studies and conclude in Section~\ref{sec:summary}.

\section{Optical model}
\label{sec:opModel}

\subsection{Model overview}

\begin{figure}[!ht]
\centering
\includegraphics[width=9cm]{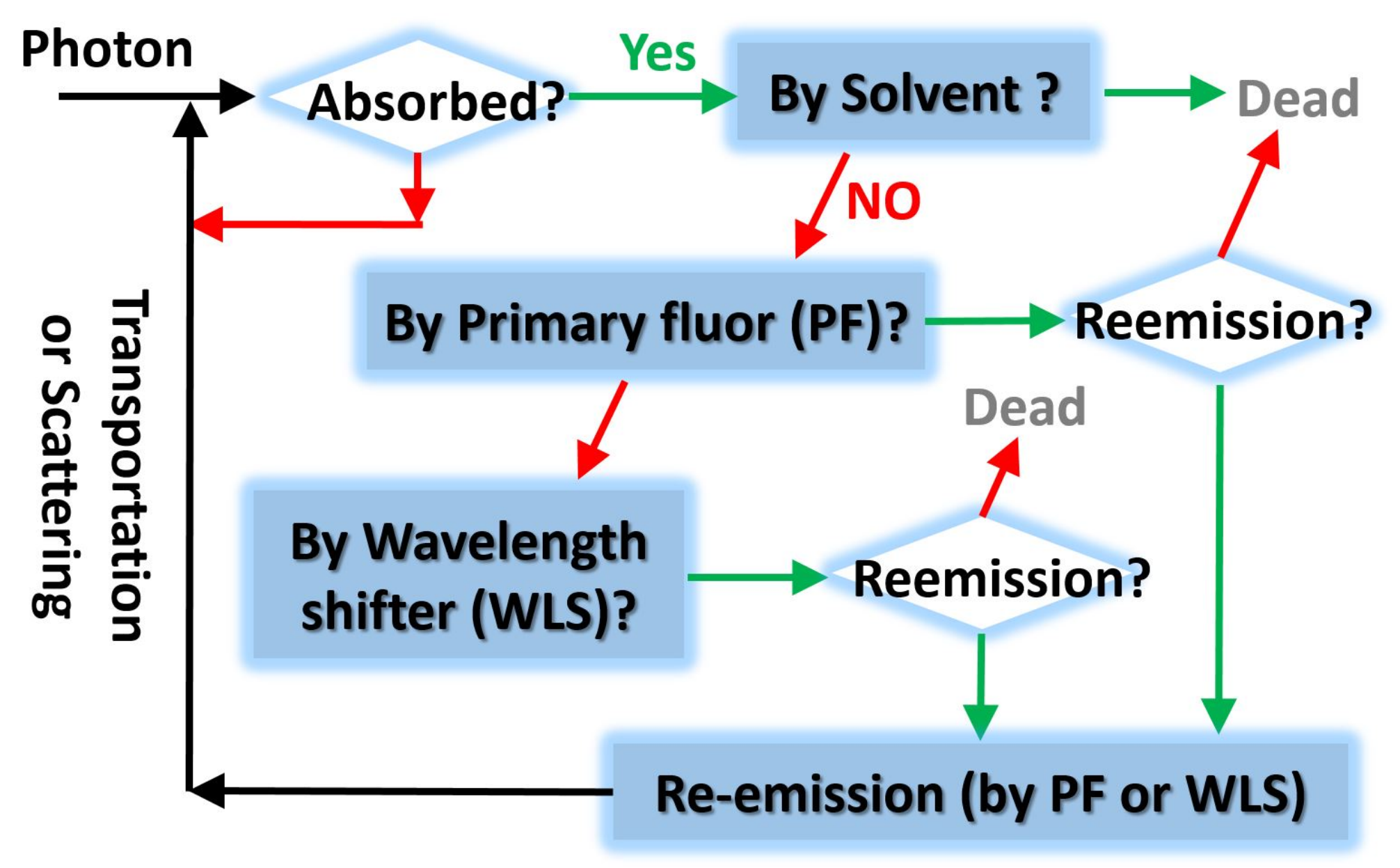}
\caption{The schematic diagram of light propagation in the new optical model.}
\label{fig:process}
\end{figure}

When an optical photon propagates in LS, there exists a competition for light absorption and scattering between all components of the LS. Once a photon is absorbed by a fluorescent molecule, a new photon may be re-emitted with a certain probability. These processes will repeat until the photon is absorbed by a molecule and no new photon is produced. The diagram in Fig.~\ref{fig:process} schematically describes the model of light propagation in a step-wise way:
\begin{itemize}
  \item If a photon is absorbed by a solvent molecule, it vanishes. However, if it is a UV photon, the solvent may still originate a re-emission, see discussions in Sec.~\ref{sec:opPar:absorption} and Sec.~\ref{sec:flourQE}.
  \item If a photon is absorbed by a primary fluor or wavelength shifter molecule, it vanishes. A new photon may be emitted with a probability corresponding to the wavelength-dependent fluorescence quantum yield of the primary fluor or wavelength shifter, respectively.
  \item If a photon is not absorbed by any component of LS, it continues to propagate.
  \item If photon scattering (Rayleigh) happens, the photon changes its direction according to the scattering cross section and continues propagation.
\end{itemize}
This refined optical model requires a deep understanding of the light propagation among LS molecules and comprehensive measurements of the optical properties. A similar and simplified modeling approach for a binary scintillator was presented in Ref.~\cite{Alimonti:2000wj}. In the following, we summarize the preparative measurements that form the basis for our more detailed optical model. In the past, the Monte Carlo simulations of many LS detectors ignored the competition of the LS components for Scattering and absorption, instead describing the complete LS as a single entity. Part of the reason for this simplification was that the measurement of all relevant optical properties is challenging. Furthermore, the simplified model does not allow to directly plug in laboratory measurements. For the LS with the same components, the simplified model needs non-trivial tuning for each different LS composition to effectively model the iterations of absorption and re-emission among the fluors, while the new optical model is able to use one set of the measured parameters of the LS components to model any LS composition.

\subsection{Optical parameters}

The key parameters of this optical model include the emission spectrum and quantum yield of all fluorescent materials, the absorption spectra of all components of the LS, as well as the Rayleigh scattering coefficient. In the following, we take the LS composed of LAB, PPO and bis-MSB as an example and review all optical parameters.

\subsubsection{Emission spectrum}

\begin{figure}[!ht]
\centering
\includegraphics[width=10cm]{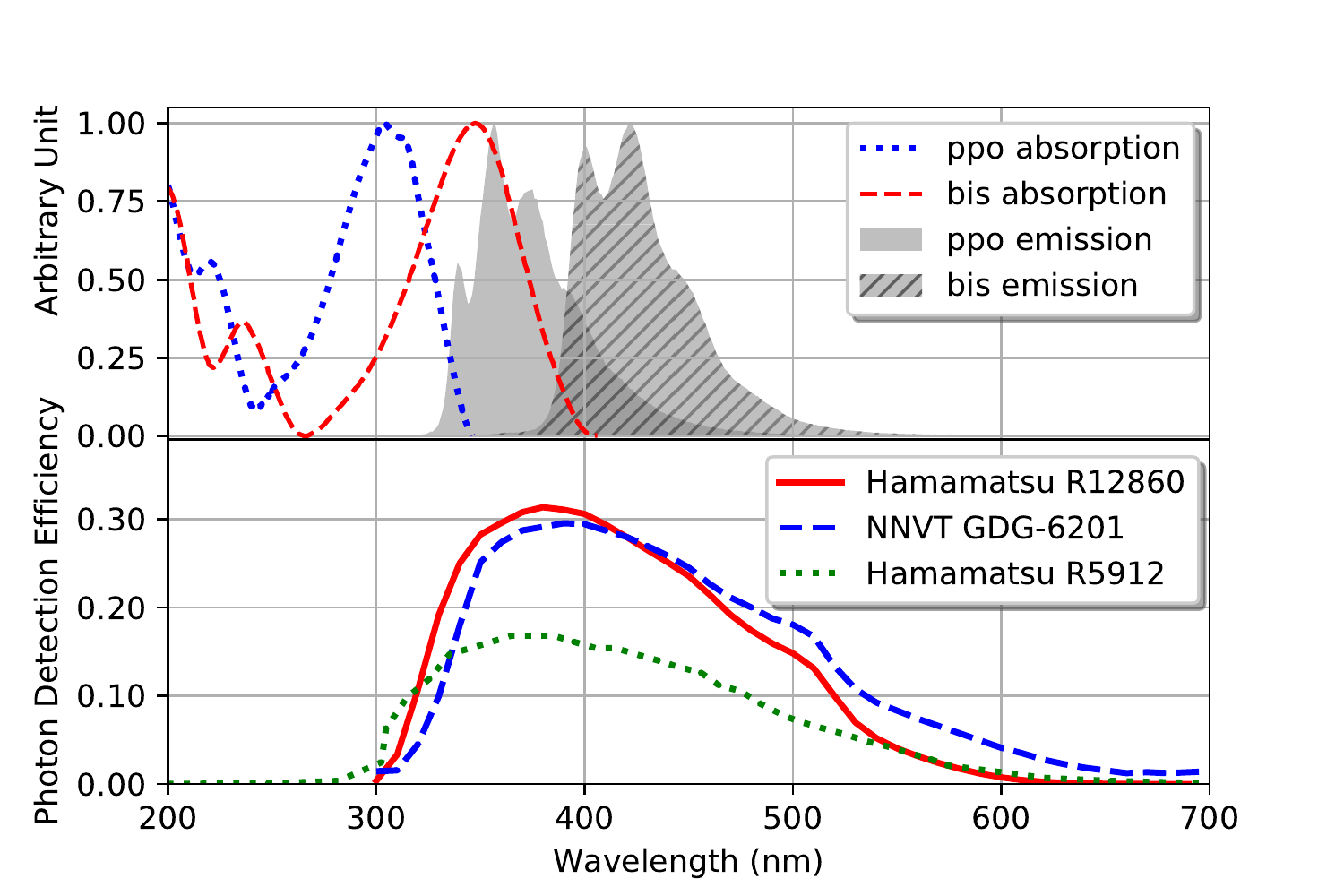}
\caption{Top: Normalized emission spectra and absorption coefficients for PPO and bis-MSB, respectively. Bottom: Typical quantum efficiency spectra of three types of PMTs.}
\label{fig:exti_emi}
\end{figure}

It is well known that, for both PPO and bis-MSB, the emission spectra and absorption spectra somewhat overlap giving rise to the self-absorption effects. In Fig.~\ref{fig:exti_emi}, the emission spectra measured with a Fluorolog Tau-3 spectrometer are shown as shaded regions, while the absorption curves are taken from~\cite{LaserDyes} for purpose of demonstration. We have as well re-measured the absorption spectra, and details are described in the next subsection.

Once an ionization event occurs within LS and excites mostly LAB molecules and provided the average distances between the LAB and PPO molecules are close enough, the de-excitation of the LAB molecules is dominantly via fast dipole-dipole interaction with the PPO molecules. As a result, the primary photons are emitted from PPO given the PPO content is at the level of a few gram-per-liter. This has been verified by measurements of scintillation decay time~\cite{Xiao2010}. Thus, the emission spectrum of PPO is used to generate the primary photons in our optical model.

It should be noted that the spectral dependence of the quantum efficiency (QE) of photomultiplier tubes (PMTs) will affect the optimal choice of LS recipe, because one would aim to have maximum light emission and detection close to the maximum of PMT QE curves. The bottom panel of Fig.~\ref{fig:exti_emi} shows the typical QE spectra of three types of PMTs: the 20-inch MCP-PMT~\cite{Wang:2012rt,NeuTel2017:Wang,Wen:2019sik} (GDG-6201 model), the 20-inch Hamamatsu R12860 PMT and the 8-inch Hamamatsu R5912 PMT. In this demonstration plot, the photon detection efficiencies at 420 nm for both MCP-PMT and Hamamatsu 20-inch PMTs are normalized to 28\%, according to recent test results from the JUNO experiment~\cite{Taup19:Zhang}. For the R5912 type, the peak efficiency is the product of the nominal quantum efficiency of 24\% and the nominal collection efficiency of 70\%. Their impact on selecting the LS recipe is investigated in Sec.~\ref{sec:modelApp}. Note that the MCP-PMT's QE will be further improved and its spectral shape may change due to the fine tuning of photocathode technology.

\subsubsection{Absorption probability of each component}
\label{sec:opPar:absorption}

The absorbance of each LS component $A_i(\lambda)$, was measured with a Shimadzu UV2550 UV-Vis spectrometer and can be expressed in terms of the molar extinction coefficient $\varepsilon_i(\lambda) = A_i(\lambda)/(\eta_i\cdot l)$, where $\lambda$ is the photon wavelength, $l$ is the light path of the quartz cuvette, $\eta_i$ is the molar concentration in mol$\cdot$L$^{-1}$ unit and $\varepsilon_i$ is in cm$^{-1}$/(mol$\cdot$L$^{-1}$) unit. The systematic uncertainties take into account~\cite{Wen2010,Feng:2015tka} the corrections from air-quartz surface reflections, fluorescence due to re-emission along the light path,
imperfect focusing of the light beam, light scattering effect, etc. To minimize the uncertainties, a set of quartz cuvettes with different light paths (2 mm, 4 mm, 10 mm and 10 cm) and a set of LS samples with different fluor concentrations were used to measure and unify the absorption in different wavelength ranges. For the extreme UV region, the absorption curve was measured with a very thin liquid layer sandwiched by two quartz sheets featuring high transparency and uniform thickness.

Due to short light path, the UV-Vis measurement has large uncertainties if the absorption length is greater than 10 m. Thus, a facility with a 1-m long tube was set up to measure the attenuation length at 430 nm~\cite{Gao:2013pua}, and the measured attenuation length was used to scale the spectral shape from the UV-Vis measurements, to obtain the absorption spectra in the full wavelength range. It should be pointed out that, in the VIS waveband the measured attenuation length consists of the Rayleigh scattering length and the absorption length and can be described as $\frac{1}{L_{att}(\lambda)} = \frac{1}{L_{Rayleigh}(\lambda)} + \frac{1}{L_{abs}(\lambda)}$. In order to obtain the absorption curve ($L_{abs}$), the Rayleigh scattering effect ($L_{Rayleigh}$) should be subtracted from the measured attenuation ($L_{att}$). Recently the $L_{Rayleigh}$ of LAB was measured to be 27.0 m at 430 nm~\cite{Zhou:2015gwa}, and it is used in this work.

\begin{figure}[!ht]
\centering
\includegraphics[width=10cm]{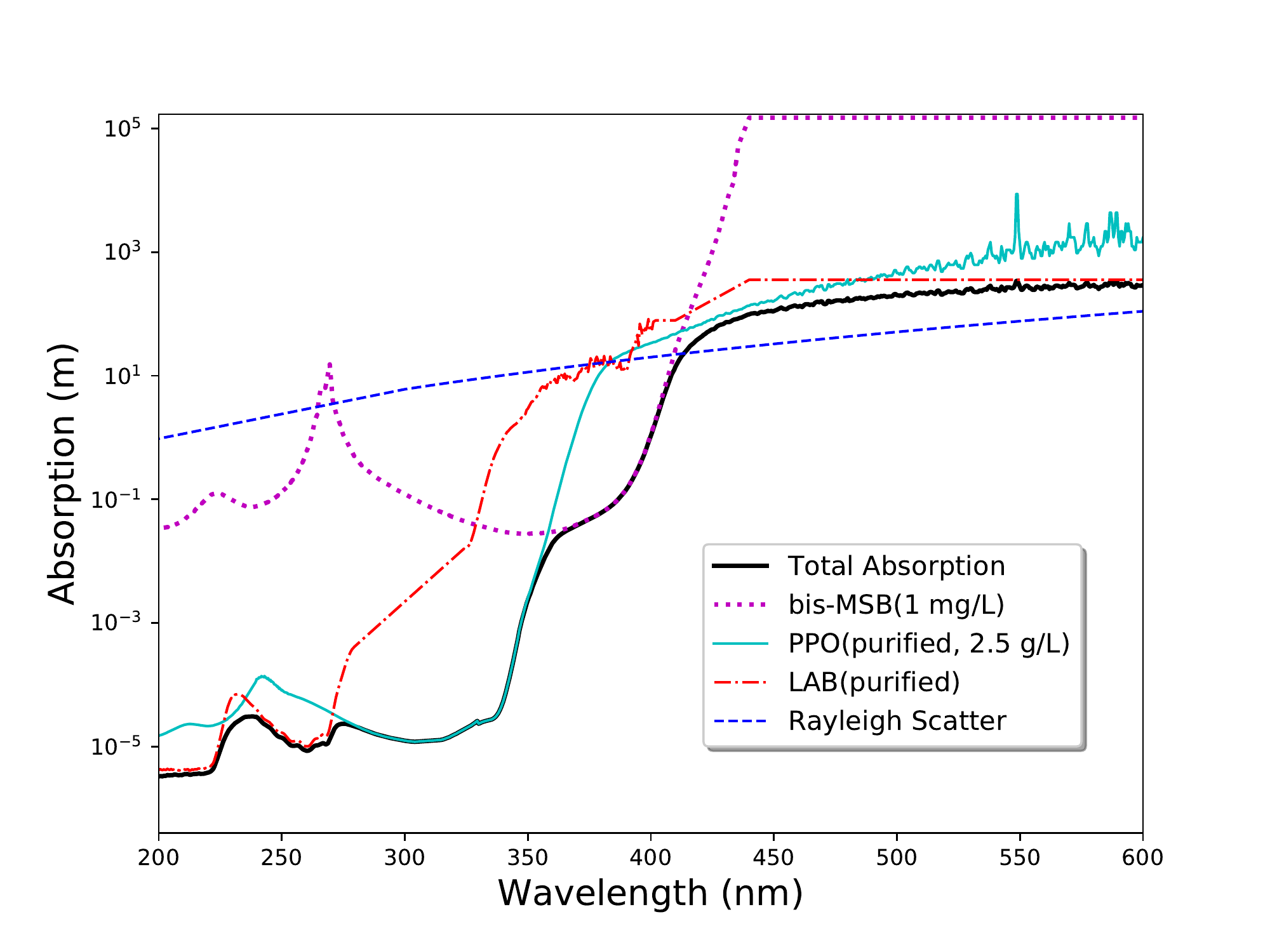}
\caption{(colour online) Measured absorption length of each LS component and total absorption, superimposed with Rayleigh scattering length.}
\label{fig:absL}
\end{figure}

Taking the LS mixture composed of LAB, PPO (2.5 g/L) and bis-MSB (1 mg/L) as an example, Fig.~\ref{fig:absL} shows the measured absorption curve of all components and the total absorption, as well as the parameterized curve of the Rayleigh scattering length in which the value at 430 nm is normalized to 27.0 m. In Fig.~\ref{fig:absL}, both LAB and PPO are purified to obtain optimal transparency. Above about 410 nm, the scattering dominates the photon transportation. When increasing the bis-MSB concentration, the steep edge between 400 nm and 420 nm in the total absorption spectrum will shift to larger wavelength.
For visible photons with wavelengths $>$430 nm, we have no measured data to constrain the absorption lengths, however we believe that the purified LAB maintains a very good transparency. Thus, we use a flat extrapolation for LAB and bis-MSB starting from 440 nm. The UV-Vis spectrum also indicates that LAB starts to have absorption in the infrared region, where the PMT photocathode is insensitive.
The transparency of LS component may affect the optimization of the LS composition. This effect is studied in Sec.~\ref{sec:modelApp}.

\begin{figure}[!ht]
\centering
\includegraphics[width=10cm]{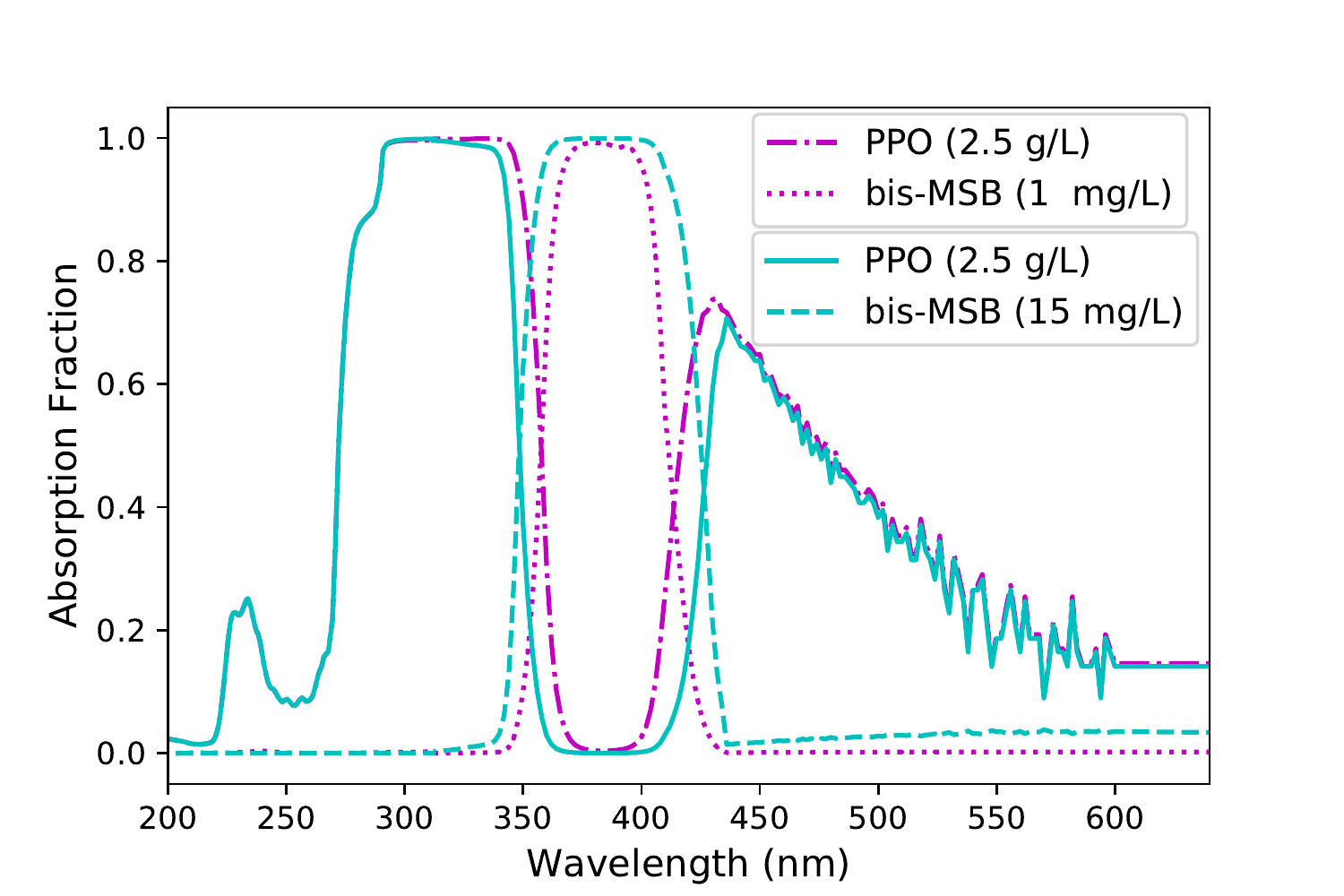}
\caption{The absorption fraction of PPO and bis-MSB if a photon is
absorbed by the LS mixture, where the PPO is 2.5 g/L
and the bis-MSB is 15 mg/L or 1 mg/L, respectively. }
\label{fig:diffAbs}
\end{figure}

In a homogeneous and stable LS mixture, if the light absorption by the solvent and the fluorescent molecules behave independently, the total absorbance is a linear combination of the absorbance of each component. This empirical feature is known as the Beer-Lambert law, which can be described as $A_{total}(\lambda) = \varepsilon_{total}(\lambda) \cdot c_{total} \cdot l = \sum \varepsilon_i (\lambda) \cdot c_i \cdot l$, where $\varepsilon_i(\lambda)$ is the molar extinction coefficient of the $i$-th component, and $A_{total}$ and $\varepsilon_{total}$ are the total absorbance and molar extinction coefficient of the LS mixture, respectively. Thus, if a photon is absorbed by the LS mixture, the absorption fraction of the $i$-th component can be calculated as $\varepsilon_i(\lambda)\cdot c_i/(\varepsilon_{total}\cdot c_{total}$). The measurements with different LS recipes agreed well with the Beer-Lambert law. An example for absorption fractions of PPO and bis-MSB is shown in Fig.~\ref{fig:diffAbs}, which is calculated with the data in Fig.~\ref{fig:absL}. It is worth to note that:
\begin{itemize}
\item Below $\sim$290 nm, LAB dominates the absorption probability. In this UV waveband, the excited LAB molecules transfer the energy to PPO molecules and the latter de-excite by emitting a photon according to PPO's quantum yield. The optical model takes this into account.

\item PPO and bis-MSB compete in absorption in two spectal regions, from $\sim$340 nm to $\sim$370 nm and from $\sim$400 nm to $\sim$430 nm, while bis-MSB dominates the absorption in-between. The higher the concentration of bis-MSB, the wider the range in which it dominates the absorption.

\item Beyond 440 nm, the absorption spectra derived from the UV-Vis measurement have large uncertainties, and there are hints that PPO still has a significant contribution to the absorption. However, the quantum yield measurement indicates that PPO has almost no quantum yield at such long wavelengths, thus re-emission is negligible in this spectral region.
\end{itemize}

\subsubsection{Fluorescence quantum yield of fluors}
\label{sec:flourQE}

The excited fluorescent molecules return to the ground state predominately via two processes: radiative de-excitation and non-radiative de-excitation. The fluorescence quantum yield corresponds to the probability of radiative de-excitation under emission of fluorescence photons~\cite{Birks1968}. According to Fig.~\ref{fig:diffAbs}, it is important to measure the fluorescence quantum yield below 370 nm for PPO and between 340 nm and 430 nm for bis-MSB, respectively.
\begin{figure}[!htb]
\centering
\includegraphics[width=11cm]{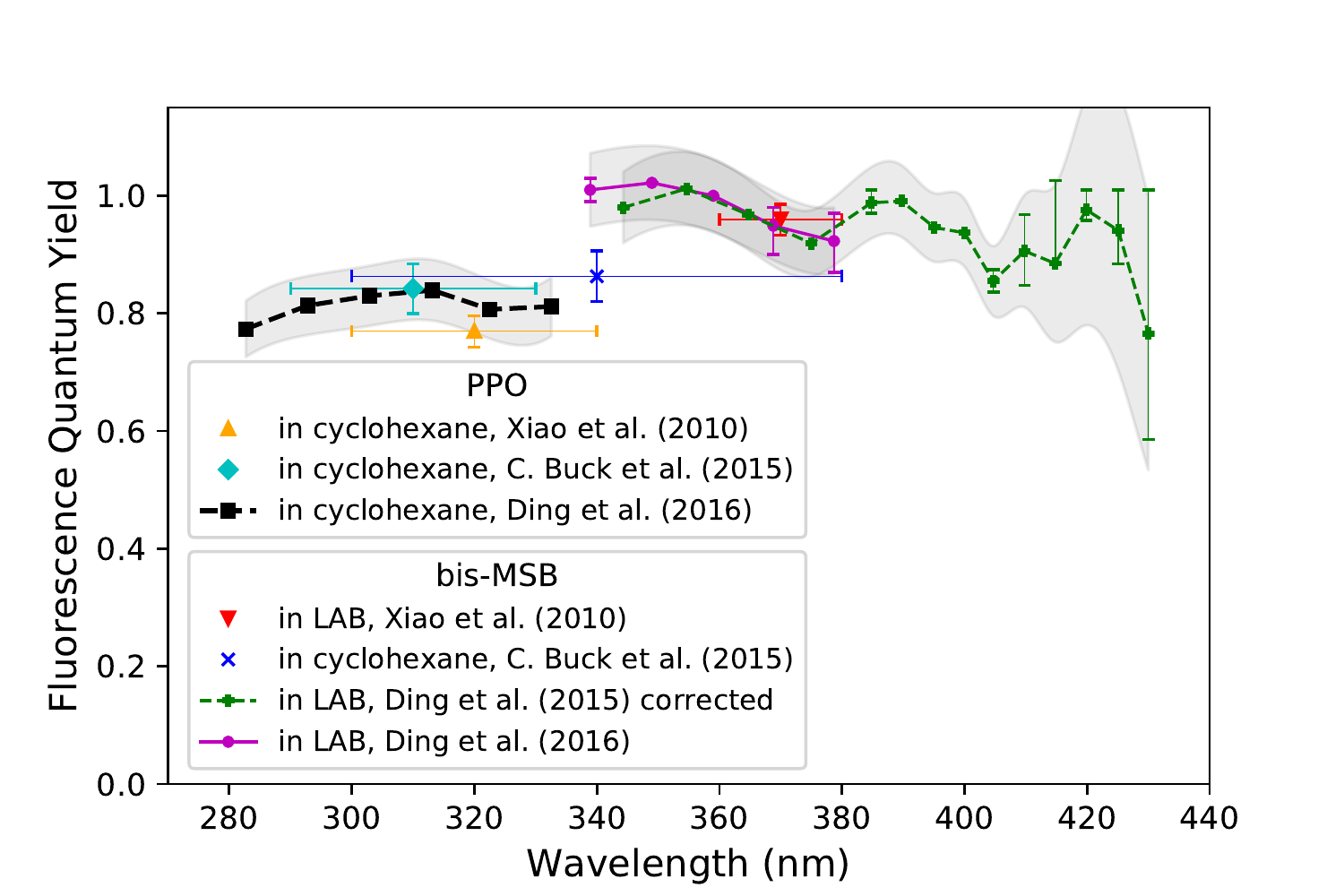}
\caption{Summary of previously measured fluorescence quantum yield of PPO and bis-MSB.}
\label{fig:ppo_bisMSB_QY}
\end{figure}

The results of previous measurements, e.g., see Ref.~\cite{Xiao2010,Feng:2015tka,Buck:2015jxa}, are summarized in Fig.~\ref{fig:ppo_bisMSB_QY}. Some comments about these measurements are listed below.
\begin{itemize}
  \item In all cases, the combination of a fluorescence spectrometer and a UV-Vis spectrometer was used. This method has intrinsic difficulties because several corrections and effects need be considered. Thus, the measured fluorescence quantum yields of PPO and bis-MSB have relative large uncertainties.

  \item It should be pointed out that the quenching effect due to the dissolution of O$_2$ in the samples can lead to an underestimated quantum yield. This effect was not considered in Ref.~\cite{Feng:2015tka}. In later measurements, this effect was reduced by sufficient N$_2$ purging during the measurements of both the absorbance spectrum and the fluorescence intensity. The re-measured quantum yield of bis-MSB was found to be about 7.5\% higher than the reported values in Ref.~\cite{Feng:2015tka}. Thus, in Fig.~\ref{fig:ppo_bisMSB_QY}, the values from Ref.~\cite{Feng:2015tka} are scaled by a factor of 1.075. Due to the correction, the bis-MSB measurements with LAB as solvent are consistent, and their weighted average value is used in our optical model, which is 0.95.

  \item For all PPO measurements in this plot, the solvent is cyclohexane, and the measurements are roughly consistent within errors.

  \item The use of different solvents may have an affect on the measured quantum yield. The bis-MSB measurement with cyclohexane as a solvent in Ref.~\cite{Buck:2015jxa} is about 10\% lower than that with LAB in Ref.~\cite{Xiao2010,Feng:2015tka}. Thus, we take the weighted average of measured PPO quantum yields in cyclohexane by Ding {\it et al.} (2016)   and scale it by a factor of 1.10, assuming that the solvent effect for PPO is similar to that of bis-MSB. The final value for PPO in LAB is 0.898 in our model.
\end{itemize}

\begin{figure}[!htb]
\centering
\includegraphics[width=10cm]{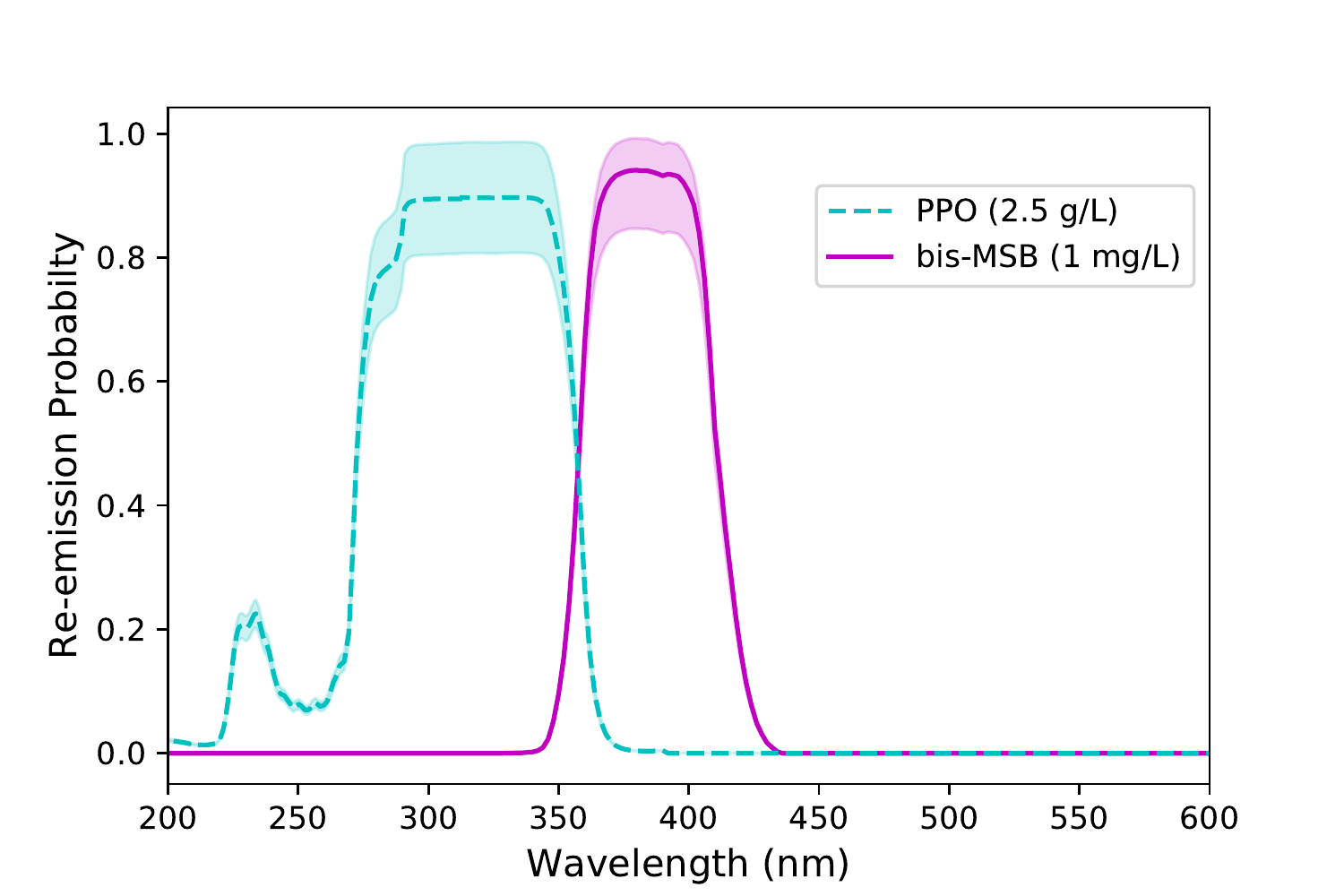}
\caption{Re-emission probability on each fluor, i.e., PPO or bis-MSB, in the LS mixture that consists of LAB, 2.5 g/L PPO and 1 mg/L bis-MSB.}
\label{fig:reemProb}
\end{figure}

For each fluor, we define the re-emission probability as the product of absorption fraction and fluorescence quantum yield, and this quantity represents the overall probability that the re-emission occurs via a certain fluor after a photon is absorbed in the LS mixture. Using the data in Fig.~\ref{fig:diffAbs} and Fig.~\ref{fig:ppo_bisMSB_QY}, the re-emission probability is calculated for PPO and bis-MSB, respectively, shown in Fig.~\ref{fig:reemProb}. It should be pointed out that our model requires the energy of re-emitted photons to be lower than that of the absorbed photon, to conserve energy in the transition.

\subsubsection{LS time profile}

The timing properties of LAB-based liquid scintillator were extensively studied through ultraviolet and ionization excitations, e.g., see Ref.~\cite{Li2011,Lombardi:2013nla}. In~\cite{Li2011}, under ultraviolet excitation, the intrinsic decay times of PPO, bis-MSB and LAB were measured to be $\sim$1.6 ns, $\sim$1.5 ns and $\sim$48 ns, respectively, and they are used to model the time delay once the re-emission occurs.

In the case of ionization by a charged particle, molecular excitations occur predominately on LAB molecules due to their dominant mass fraction. Based on the measurements of the emission spectra, we concluded that energy transfer from LAB to PPO molecules is efficient for concentrations of few g/L, and the primary fluorescent photons are emitted by PPO. However, the energy transfer from LAB to PPO causes a certain delay in the scintillation times. The data in Ref.~\cite{Li2011,Lombardi:2013nla} indicates that the size of the delay depends on the PPO concentration.

Moreover, the contribution of the slow scintillation components depends on the type of charged particle causing the ionization. In our optical model, the time profile of the primary scintillation photons is based on the measured profile of LAB+PPO without the addition of bis-MSB.

\section{Model Validation}
\label{sec:experiment}

\begin{figure}[htb]
\centering
\includegraphics[width=12cm]{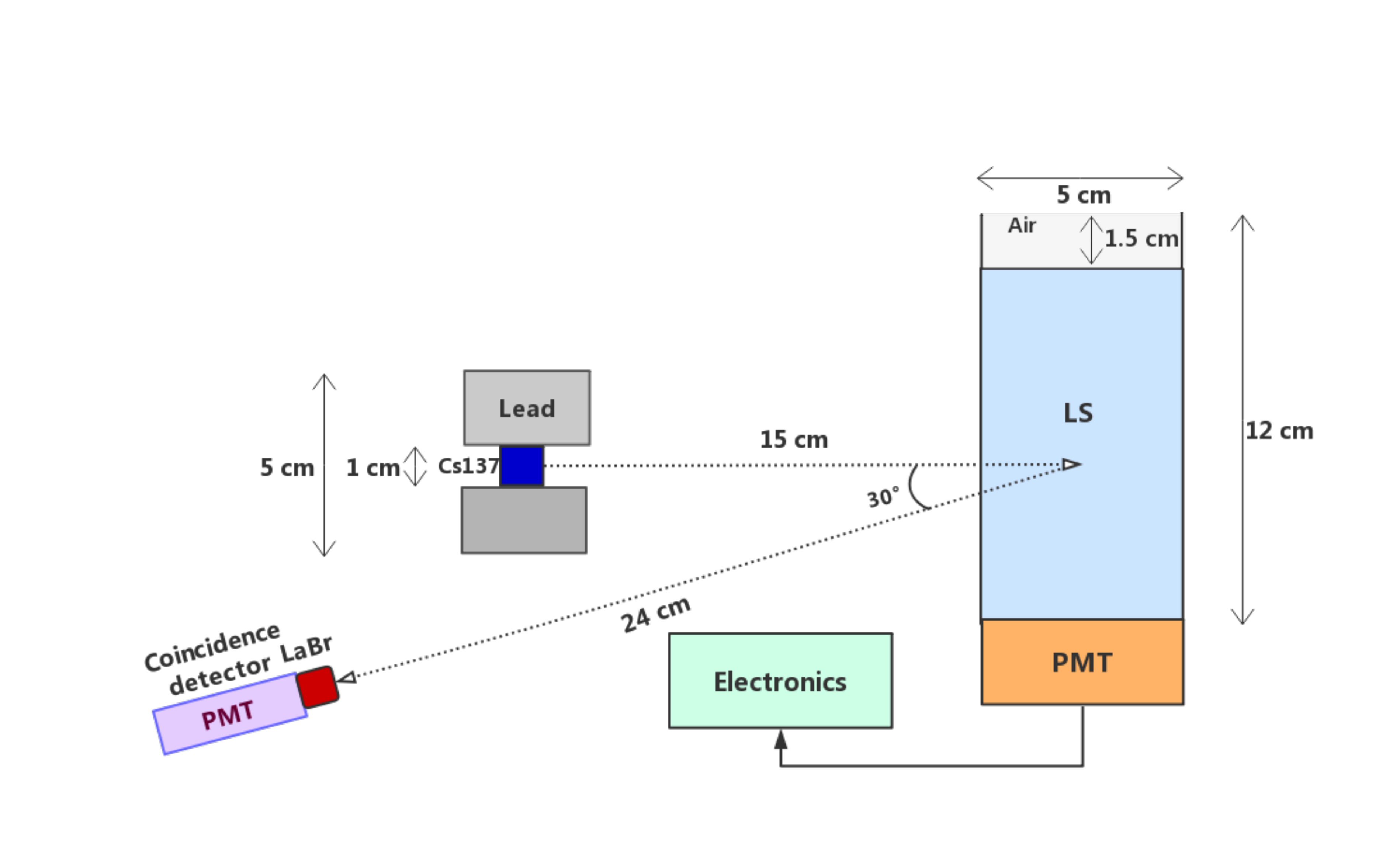}
\caption{Diagram of the bench-top experiment to validate the optical model.}
\label{fig:exp}
\end{figure}

\begin{figure}[htb]
\centering
\includegraphics[width=9cm]{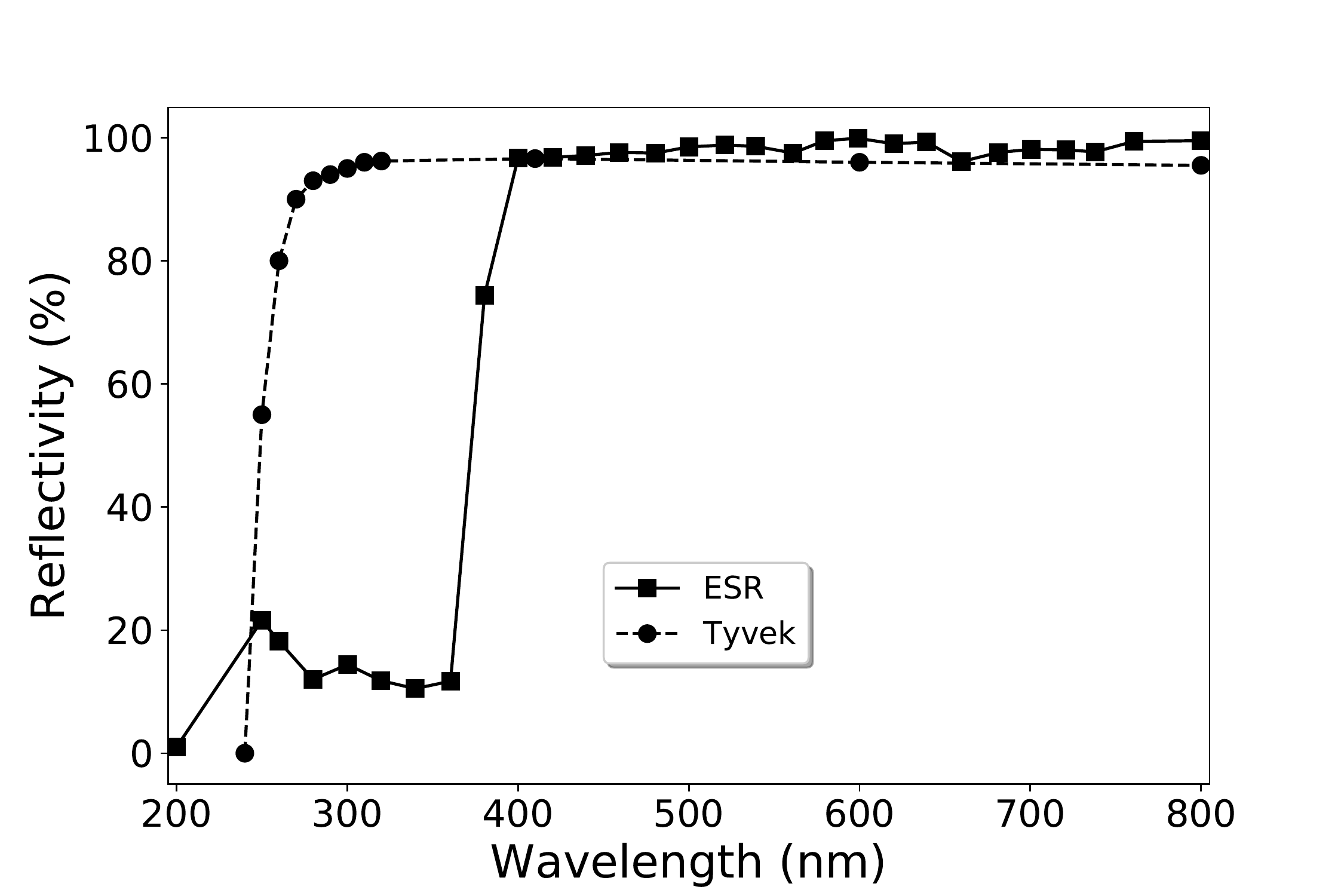}
\caption{The measured specular reflectivity of ESR and the diffuse reflectivity of the Tyvek films used in the bench-top experiment.}
\label{fig:filmRef}
\end{figure}

To validate the optical model, a bench-top experiment was set up in a dark room to avoid external light. As shown in Fig.~\ref{fig:exp}, a collimated $^{137}$Cs $\gamma$ source was deployed pointing to the center of the LS sample that contained in a cylindrical quartz vessel with a height of 12 cm and a diameter of 5 cm. A 2-inch PMT (ET9814B) was coupled at the bottom of the LS vessel to measure the light output. Two types of reflective films, ESR$^{\rm TM}$ (Vikuiti$^{\text{TM}}$ Enhanced Specular Reflector Film) and Tyvek$^{\rm TM}$, were used to cover the top and wrap the barrel surfaces of the vessel to increase the light collection. The measured reflectivity of ESR can be found in~\cite{An:2015qga}. The Tyvek film we used is a highly reflective multi-layer film formed from two pieces of 1082D Tyvek bonded with a layer of polyethylene~\cite{Dayabay:2014vka}, and the measured reflectivity in air is more than 96\% and almost constant for wavelengths from about 300 to 800 nm. Fig.~\ref{fig:filmRef} shows the measured reflectivity curves of the ESR and the Tyvek films. A cylindrical Lanthanum-Bromide crystal was deployed at a fixed angle with respect to the incident $\gamma$ ray beam with its axis pointing to the center of the LS vessel, in order to tag the Compton scattered $\gamma$ rays. The LS scintillation signal was recorded only if a coincident signal from the Lanthanum-Bromide crystal was observed, so that the measured light output was corresponding to the approximately mono-energetic electrons from Compton scattering.

A set of LS samples was prepared, with the concentration of PPO fixed at 3.0 g/L, while the concentrations of bis-MSB were 0.0, 0.5, 1.0, 1.5, 2.0, 3.0, 4.0, 5.0 and 10.0 mg/L, respectively. The LS vessel was cleaned with Alconox detergent, rinsed and dried every time before changing the LS sample. Oxygen quenching was eliminated by N$_{2}$ purging of all samples prior to the measurements. The PMT waveforms were digitized and read out by a commercial 1-GHz FADC unit (CAEN DT5751). The energies of the recoil electrons were measured via the integration of the waveforms after baseline subtraction.

\begin{figure}[!ht]
\centering
\includegraphics[width=8cm]{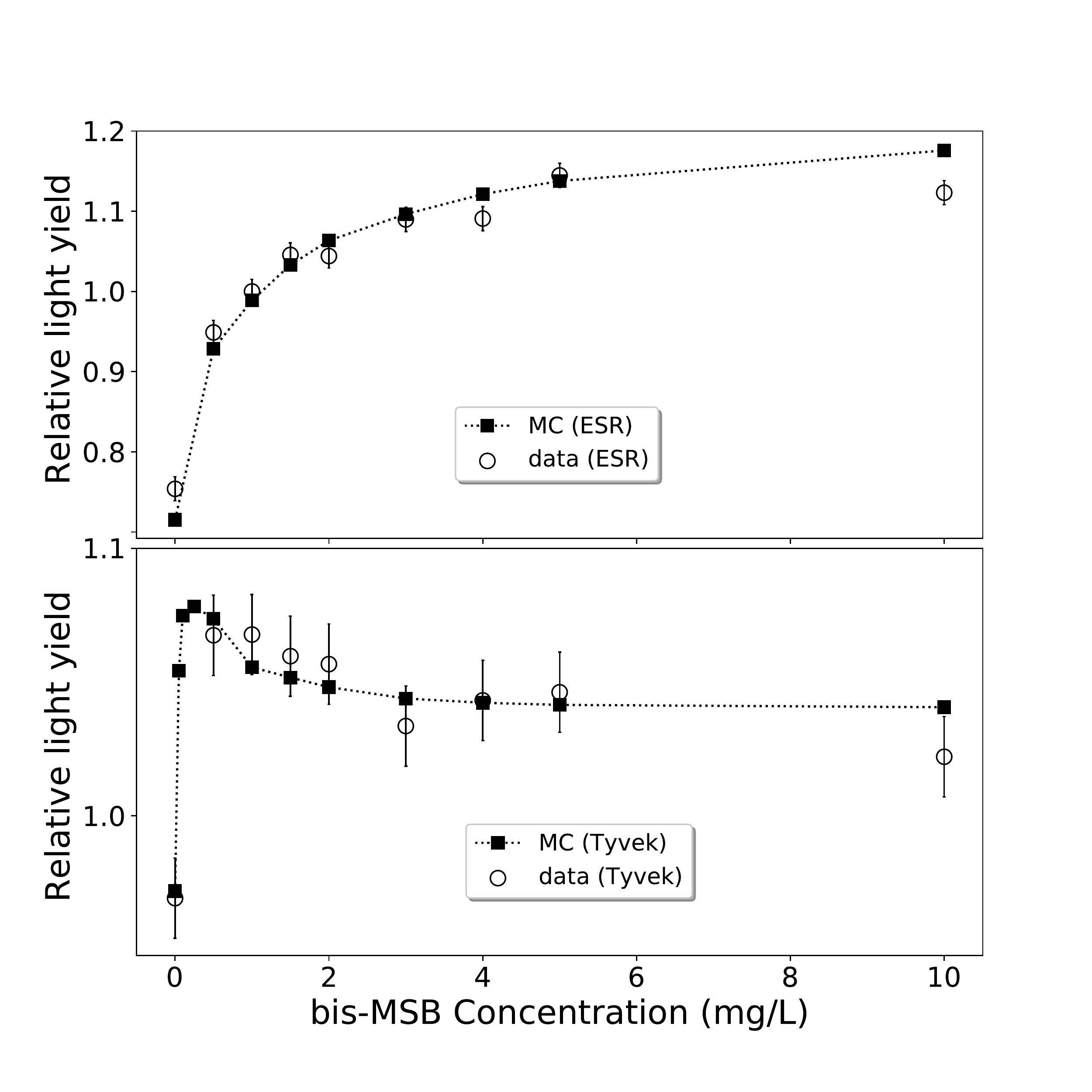}
\caption{Validation of the optical model with a bench-top experiment:
the comparison between the measured and the simulated light output as function of the concentration of bis-MSB separately for ESR and Tyvek films as reflectors. The smooth dashed curves connect the data points from simulation.}
\label{fig:varification}
\end{figure}

The measured energy spectra of the recoil electrons in each of the LS samples were fitted with the Crystal Ball function~\cite{cbfunction} to obtain the average number of photoelectrons detected.
A Geant4-based Monte Carlo (MC) simulation was developed for this bench-top experiment. The MC implemented the measured geometries of the experimental setup and the new optical model including all LS optical properties described above as well as the PMT quantum efficiency curve and the reflectivity curves of ESR and Tyvek.
Fig.~\ref{fig:varification} illustrates the comparison between the measured and the simulated light output as function of the concentration of bis-MSB separately for ESR and Tyvek films as reflectors.
This comparison is based on relative light output, and the data point at 1 mg/L bis-MSB with ESR as reflector is normalized to be 1.
The curve shapes are quite different, mainly because the reflectivity below 390 nm of the ESR film drops rapidly compared to that of the Tyvek film, as shown in Fig.~\ref{fig:filmRef}. A normalization factor $\eta$ was applied to the MC light output in order to permit the comparison with the measured data points. The factor $\eta$ was determined via minimizing the $\chi^2$ function:
\begin{equation}
\label{eq:scale}
\chi^{2} = \sum_{i} \left(\dfrac{d^{exp}_i-d^{sim}_i\cdot\eta}{\sigma^{err}_i}\right)^2,
\end{equation}
where the subscript $i$ represents different bis-MSB concentrations, $d^{exp}_i$ and $d^{sim}_i$ are the experimental data and MC data, respectively.
$\sigma^{err}_i$ is the error of $d^{exp}_i$, including both the statistical error ($<$0.4\%) and the systematic error (1.5\%) estimated by repeated tests. The systematic error is due to the PMT gain variation, imperfect wrapping of the reflective films, imperfect geometry parameters, QE non-uniformity of the PMT photocathode, etc. The error bars of the MC points are statistical only and very small.
The decent agreement between the bench-top measurements and the Monte Carlo predictions with the best-fit $\eta$, shown in Fig.~\ref{fig:varification}, provides a good validation of our model.

\section{Future application}
\label{sec:modelApp}

A straightforward application of this optical model is to optimize the LS recipe to maximize the photoelectron yield for a LS detector. To demonstrate this application, a hypothetical spherical LS detector is constructed in a Geant4-based Monte Carlo simulation. The center is a LS target with a certain radius. The PMTs are deployed at a relatively large radius and arranged in a way to have the maximum photocathode coverage, and water is filled between the LS target and PMTs. The absorption length of water is set to be infinite to eliminate the loss of photons in water. Different radii of the LS target volume (2.0 m, 6.5 m and 17.7 m) are investigated, corresponding to the detector sizes of Daya Bay, KamLAND and JUNO, respectively. The LS has LAB as a solvent, 2.5 g/L PPO as primary fluor and different concentrations of bis-MSB (0 mg/L, 0.1 mg/L, 0.5 mg/L, 1 mg/L, 4 mg/L and 7 mg/L). Mono-energetic electrons are injected at the detector center to obtain the photoelectron yield for each detector configuration.

\begin{figure}[!ht]
\includegraphics[width=8.6cm]{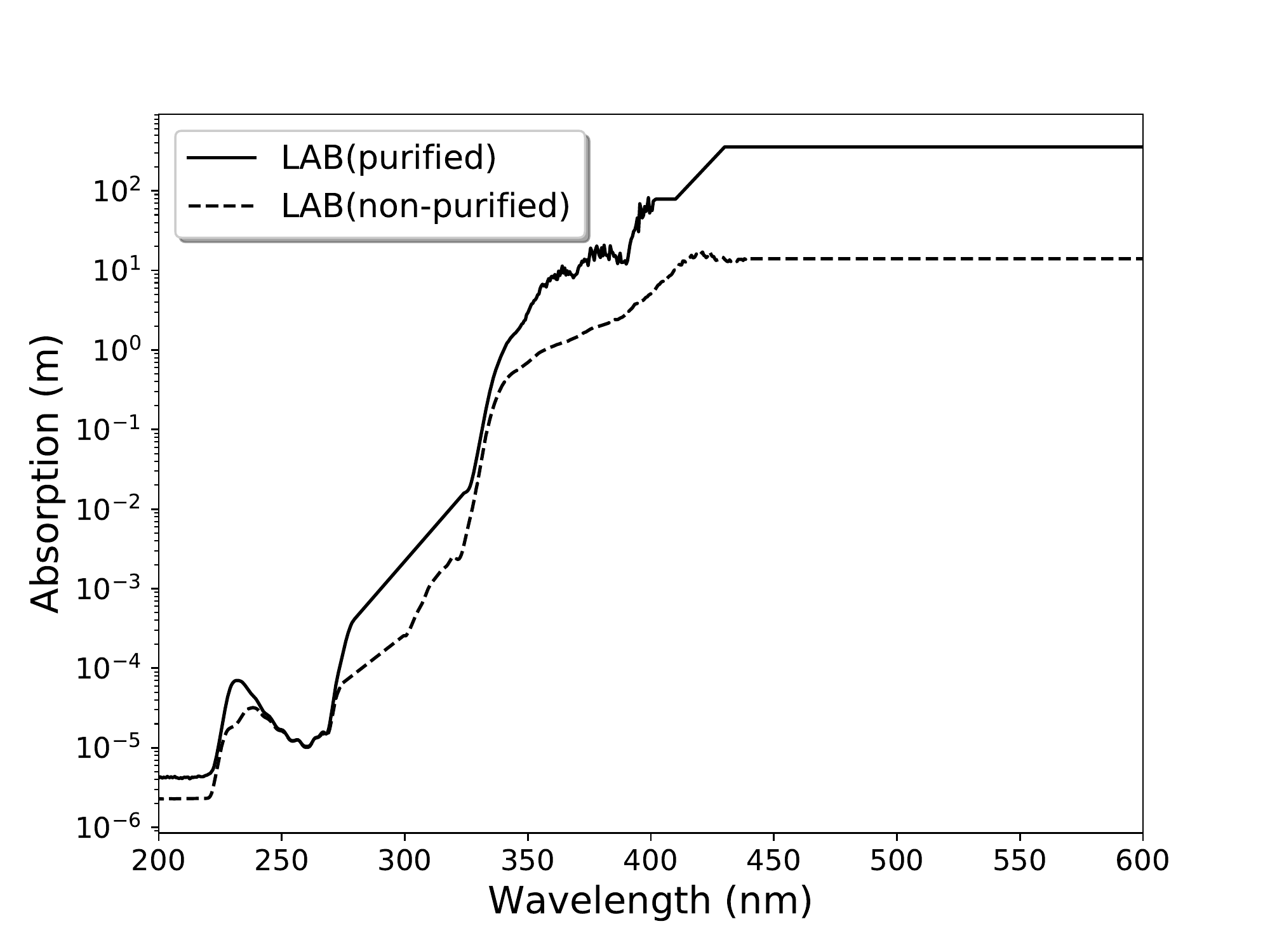}
\includegraphics[width=8.6cm]{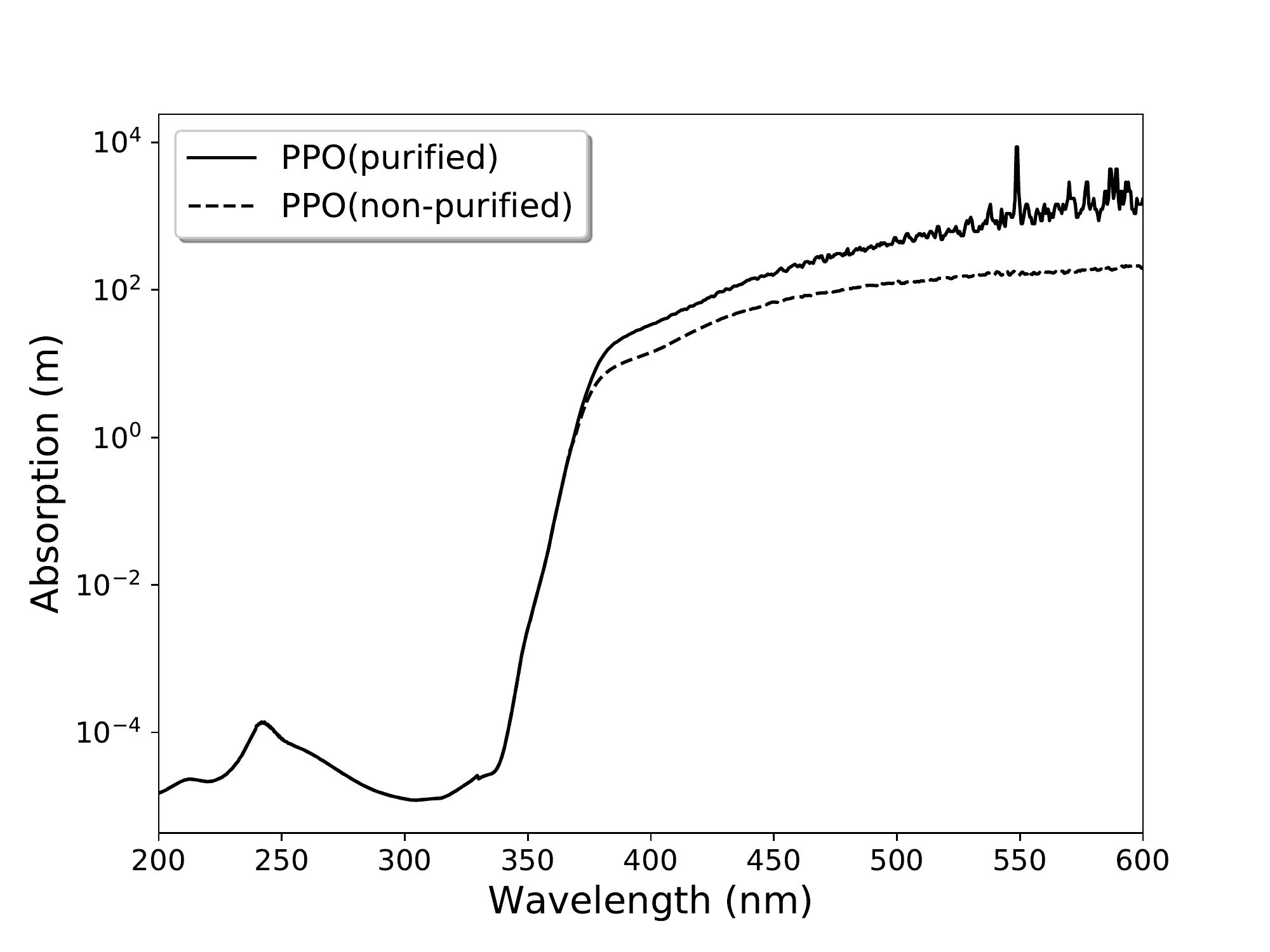}
\caption{Absorption spectra with and without purification for LAB and PPO, respectively.}
\label{fig:LABpurity}
\end{figure}

In order to investigate how the transparency of LS component affects the optimization of the LS recipe, the absorption curves of LAB and PPO with and without purification are tested, as shown in Fig.~\ref{fig:LABpurity}. The curves for purified components correspond to that expected for Al$_2$O$_3$. In addition, different spectral dependencies of the PMT quantum efficiency (shown in Fig.~\ref{fig:exti_emi}) are investigated. The main observations are summarized as below.

Fig.~\ref{fig:relYield} shows the resulting photoelectron yields for LS mixtures that have not been purified and the results for optically purified LS, respectively. For each combination of LS radius and PMT QE spectrum, the simulated light output has been normalized to that obtained at 7 mg/L bis-MSB concentration. We observe that:

\begin{figure}[!h]
\centering
\includegraphics[width=8.6cm]{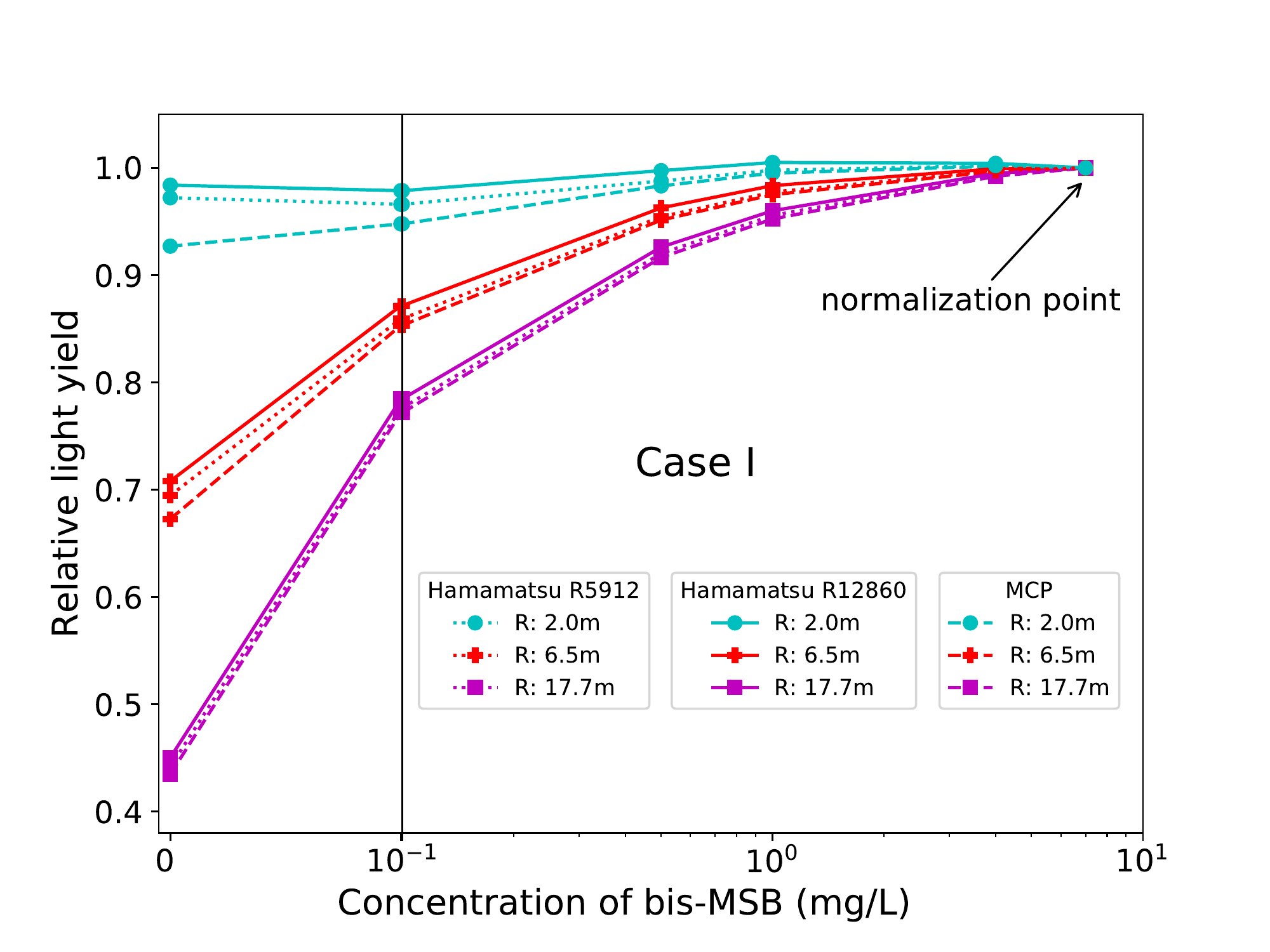}
\includegraphics[width=8.6cm]{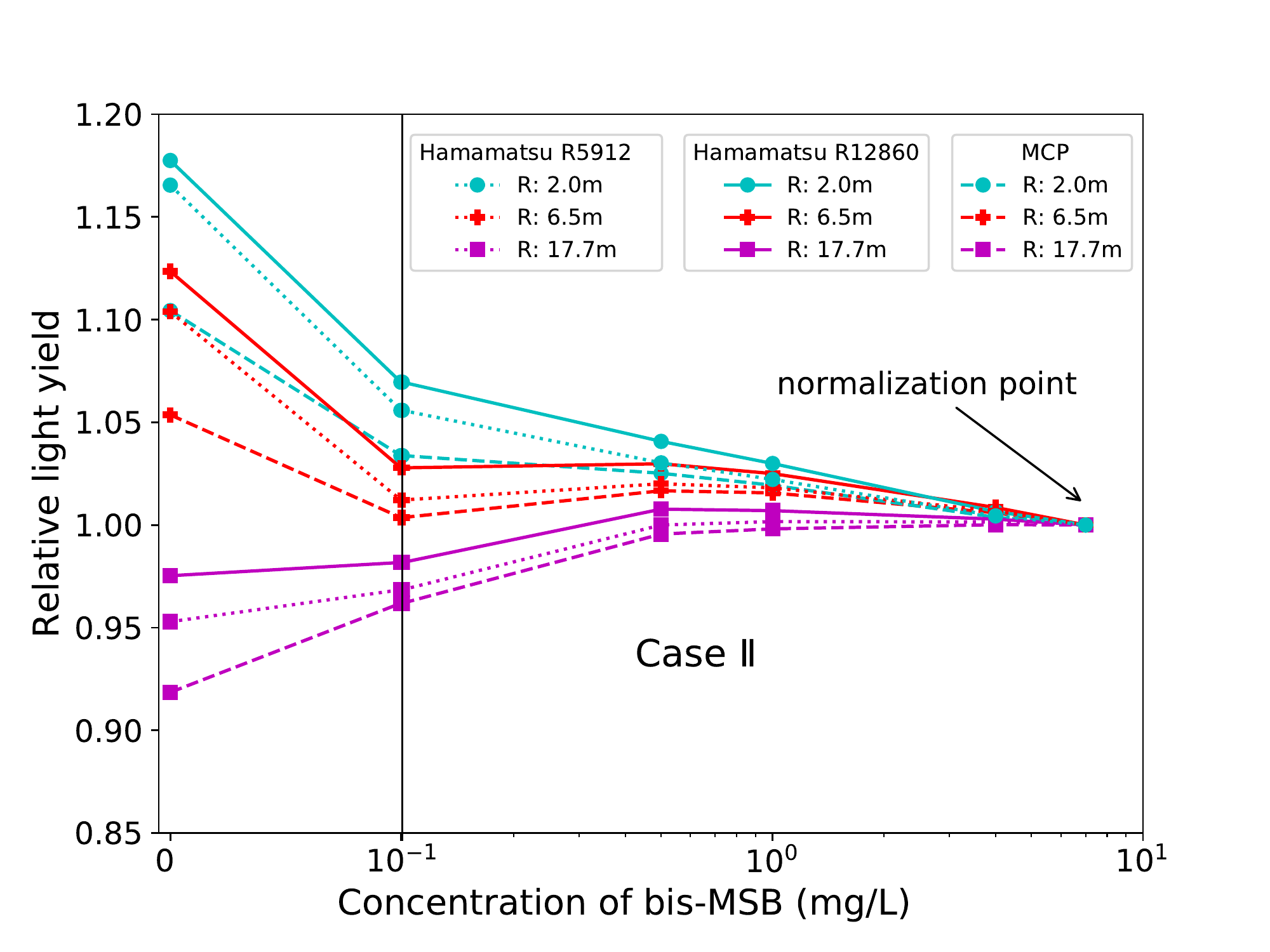}
\caption{(colour online) Relative photoelectron yields for LS mixtures that have not been purified (Case I) and the results for optically purified LS (Case II). The solid, dotted and dashed curves correspond to using the QE spectra of R12860, R5912 and MCP-PMT, respectively. For each curve, the light output is normalized by that obtained at 7 mg/L bis-MSB concentration.}
\label{fig:relYield}
\end{figure}

\begin{figure}[!htp]
\centering
\includegraphics[width=12cm]{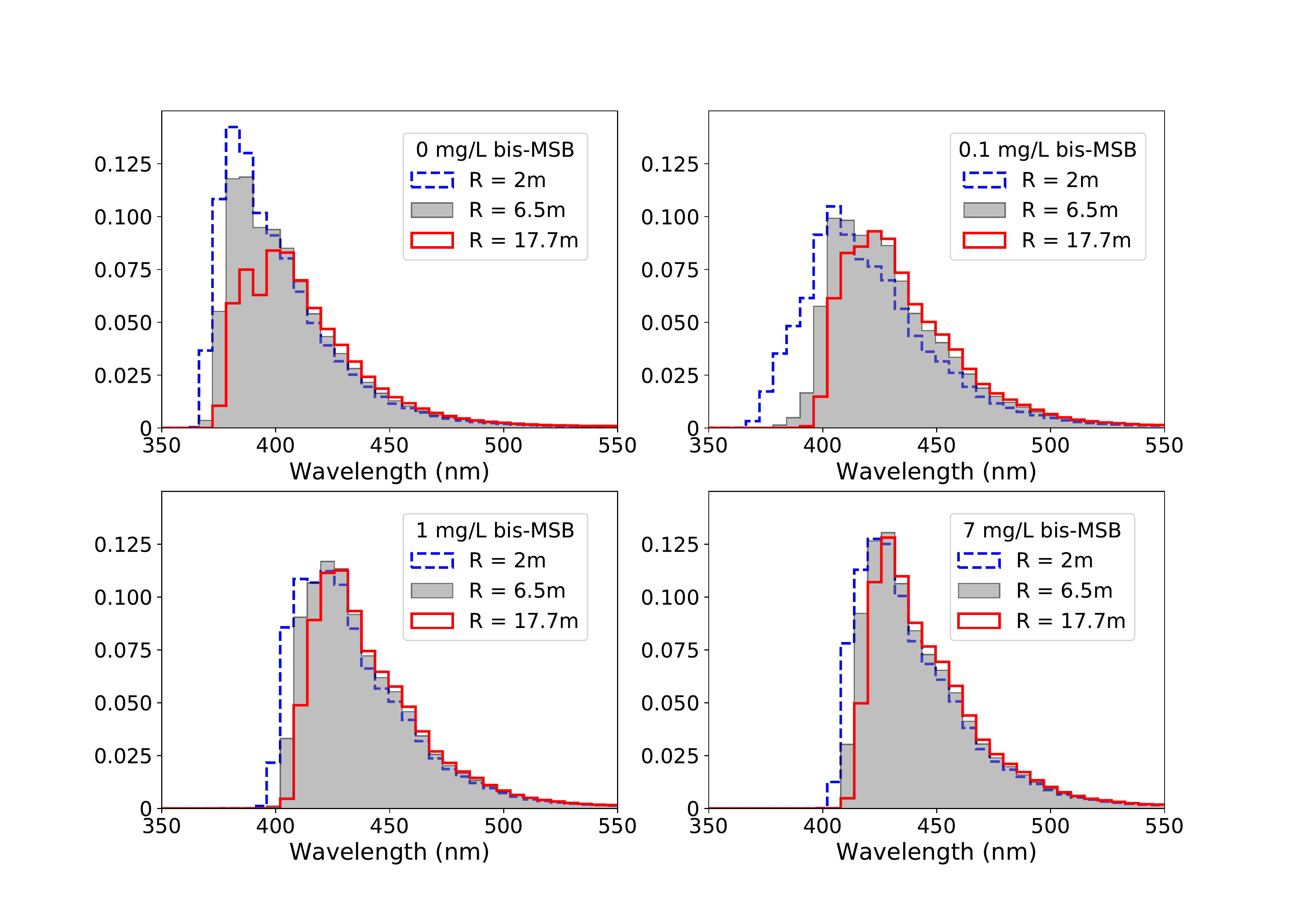}
\caption{Wavelength distributions of the photons arrived at PMTs in MC simulation, for the case that the LS is optically purified. A normalization factor is applied to all distributions, where the area of the distribution under 7 mg/L bis-MSB and $R=2$ m is normalized to be 1.}
\label{fig:photonWL}
\end{figure}

\begin{itemize}
\item The light output converges at a few mg/L bis-MSB for each configuration.
\item For the non-purified LS (Case I), bis-MSB plays a more significant role in improving the photoelectron yield as the LS radius increases.
\item For the optically purified LS (Case II) and zero content of bis-MSB , the different spectral shape of PMT QE can vary the light output by about $(5\cdots8)$\%. This can be explained by Fig.~\ref{fig:photonWL}, which illustrates the wavelength distributions of photons arrived on PMTs for different LS radii. Compared to the 7 mg/L bis-MSB case, the arrival photons in the case without bis-MSB have significant or even dominant contribution of UV photons. According to the QE spectra in Fig.~\ref{fig:exti_emi}, the larger fraction of PMT QE in UV waveband, approximately the higher the relative light output.
\item In the Case II, for small (2.0 m) or even medium (6.5 m) LS radius, the addition of bis-MSB actually decreases the light output, while for very large LS radius (17.7 m), the addition bis-SMB can still increase the light output by about $(3\cdots8)$\%.
\end{itemize}

\begin{figure}[!htp]
\includegraphics[width=9cm]{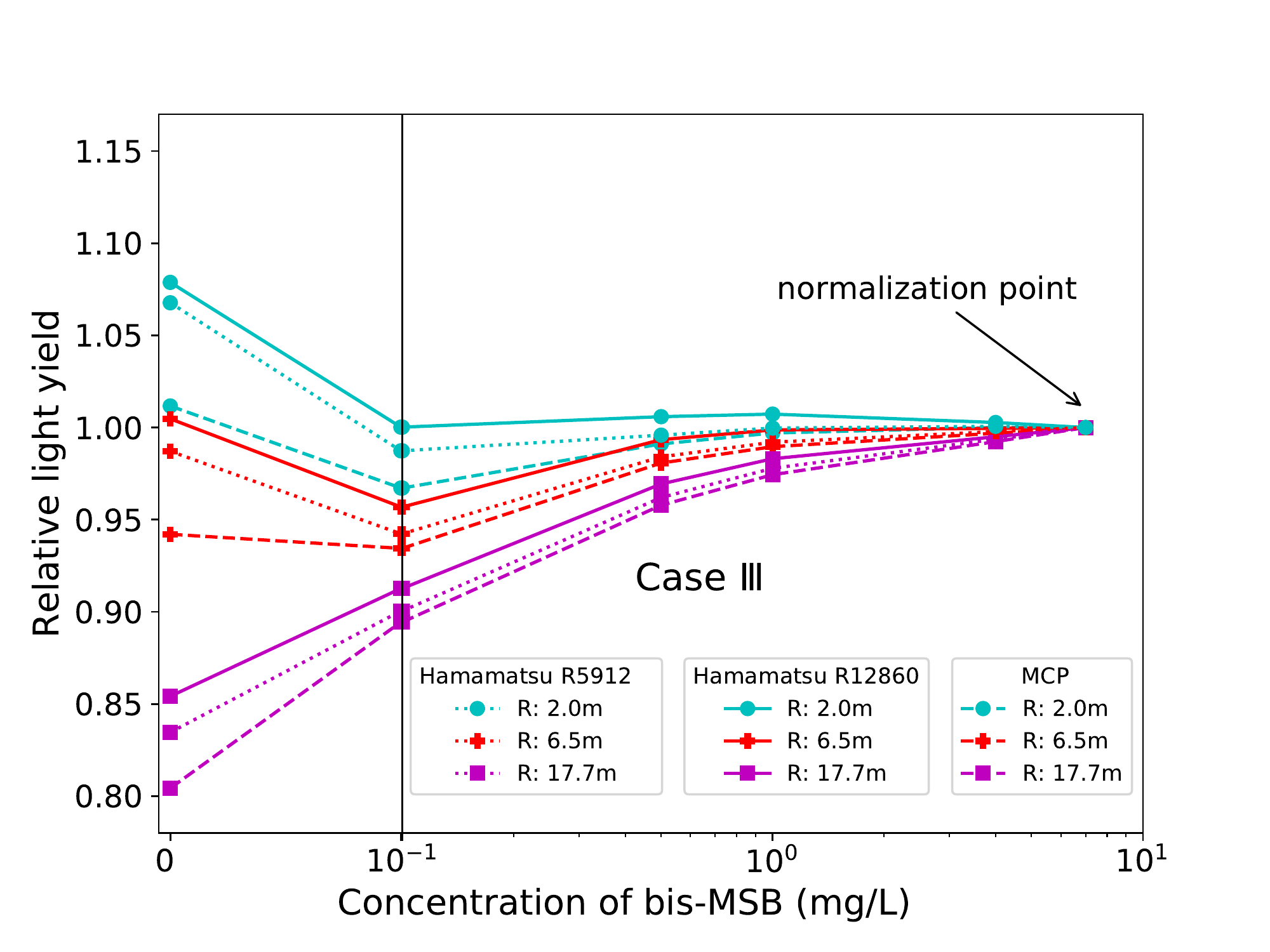}
\includegraphics[width=9cm]{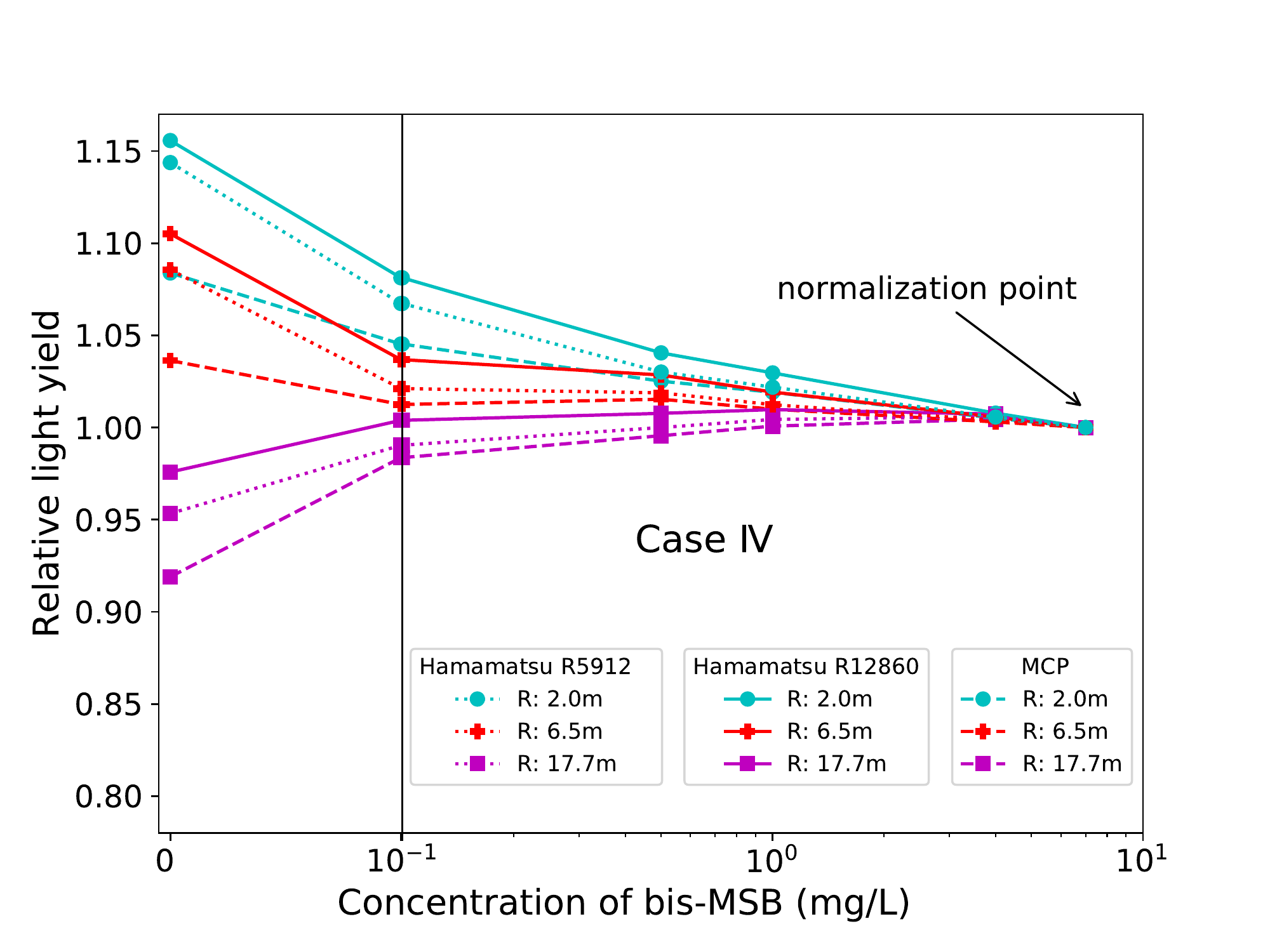}
\caption{(colour online) Same definition as Fig.~\ref{fig:relYield}. For Case III, the quantum yields of PPO and bis-MSB to be 0.8 and 1.0, respectively. For Case IV, the quantum yields of both PPO and bis-MSB are set to be 1.0.}
\label{fig:relYieldPurifiedPerfectPPOandBisMSB}
\end{figure}

It should be noted that the measured quantum yield of PPO and bis-MSB have uncertainties, and the exact values, in particular their ratios, also affect the dependency of the relative light output on the bis-MSB concentration. The quantum yields of PPO and bis-MSB used for Fig.~\ref{fig:relYield} are described in Sec.~\ref{sec:flourQE}. Thus, we investigate two other cases:
\begin{itemize}
  \item Case III: Set the quantum yields of PPO and bis-MSB to be 0.8 and 1.0, respectively.
   \item Case IV: Set the quantum yields of both PPO and bis-MSB to be 1.0.
\end{itemize}
The simulation results are shown in Fig.~\ref{fig:relYieldPurifiedPerfectPPOandBisMSB}. In both cases, the light output still converges at a few mg/L bis-MSB concentration. The curves of Case IV are similar to those of Case II. For Case III, the curves seem to rotate anti-clockwise a bit compared to those of Case II, and bis-MSB is more useful to improve light output for the large LS radius. These results indicate that, although the measured quantum yields of PPO and bis-MSB have large uncertainties, the optimal bis-MSB concentration is at a few mg/L level, which is contradictory with the choices of earlier experiments shown in Table.~\ref{tab:exp}.

In addition, for scintillators based on solvents such as pseudocumene, that have intrinsically less transparency than LAB in the VIS waveband, adding bis-MSB will improve the light output. Both KamLAND and Borexino did not adopt bis-MSB. However, the KamLAND2-Zen proposal foresees to use an LAB-based LS, most likely with PPO and bis-MSB, to improve its light yield~\cite{Azusa2018}.

\section{Conclusions}
\label{sec:summary}

In the present work, we have introduced a generic optical model for Monte Carlo simulations, which advances the handling of the competitive photon absorption and re-emission processes of the scintillator components. This model has been validated with a bench-top experiment using a relative small LS volume. This model requires good understanding and precise measurement of the optical properties of each component of LS. Moreover, we have demonstrated its capability of optimizing the LS recipe to maximize the light collection for any particular combination of detector size and PMT QE response. We expect this model to be valuable for designing future LS-based detectors, as well as improving the agreement between the Monte Carlo simulation and the experimental data in currently on-going experiments.
For example, in order to verify the design of LS purification system of the JUNO experiment, a LS pilot plant was built at the Daya Bay experimental site~\cite{Lombardi:2019epz}, and the 20-ton Gadolinium-doped LS in one anti-neutrino detector of Daya Bay was replaced with the purified LS without Gd. During the LS replacement, different sets of PPO and bis-MSB concentrations were tested in a step-wise way, and for each LS recipe the light yield was measured with $^{60}$Co source. Our model would be useful to understand this data. Results will be reported in a separate publication.

\section*{Acknowledgment}

This work was partially supported by the National Key R\&D Program of China under Grant No. 2018YFA0404100, by the Strategic Priority Research Program of the Chinese Academy of Sciences under Grant No. XDA10010100, by the National Natural Science Foundation of China under Grant No. 11575224 and by the CAS Center for Excellence in Particle Physics (CCEPP). The authors would like to thank Yayun Ding for her invaluable help on LS sample preparation and measurements, and Mengchao Liu and Boxiang Yu for their supports on the bench-top experiment, and Xiaobo Li, Hualin Xiao, Xuefeng Ding and Hangkun Xu for useful discussions during the long history of this work.

\end{document}